\documentstyle[prl,aps]{revtex}
\begin{document}
\draft
\twocolumn[\hsize\textwidth\columnwidth\hsize\csname@twocolumnfalse\endcsname
\title{Symmetries of the stochastic Burgers equation}
\author{E.V.~Ivashkevich}
\address{Dublin Institute for Advanced Studies,
10 Burlington Road, Dublin 4, Ireland and\\
Laboratory of Theoretical Physics,
Joint Institute for Nuclear Research,
Dubna 141980, Russia}
\date{\today}
\maketitle
\begin{abstract}
All Lie symmetries of the Burgers equation driven by an external random
force are found. Besides the generalized Galilean transformations,
this equation is also invariant under the time reparametrizations.
It is shown that the Gaussian distribution of a pumping force is not
invariant under the symmetries and breaks them down leading to the
nontrivial vacuum (instanton). Integration over the volume of the
symmetry groups provides the description of fluctuations around the
instanton and leads to an exactly solvable quantum mechanical
problem.
\end{abstract}
]

{\it Introduction.}---Statistical theory of turbulence was put forward by 
Kolmogorov in 1941
\cite{ANK} and has since been developed intensively. The cornerstone of
although simple but surprisingly robust Kolmogorov's dimensional analysis
is the assumption that in the fully developed turbulence there is a range
of scales where the velocity structure functions are universal; i.e.,
independent of the cutoffs provided by the scales of energy pumping and
dissipation. Much efforts have been made to understand whether there are
fluctuation corrections to the mean field scaling exponents predicted by
Kolmogorov and whether these corrections depend on the dissipation or
pumping scale \cite{MY}. Nevertheless, the problem is still far from being
solved.

Recently, the one-dimensional Burgers equation
\begin{equation}
u_t + uu_x - \nu u_{xx} = f(x,t) \label{burgers}
\end{equation}
driven by Gaussian random force $f(x,t)$ with the zero mean and covariance
\begin{equation}
\langle f(x,t)f(x',t') \rangle = \kappa(x-x') \delta (t-t'),
\label{covariance}
\end{equation}
has been in the focus of quite a number of studies. The reason is that this
equation is the simplest one that resembles the analytic structure of the
Navier-Stokes equation and, at least formally, is within the scope of
applicability of the Kolmogorov's arguments. Extensive numerical simulations
of this equation \cite{CY}, although reproducing Kolmogorov's scaling exponents
for the energy spectrum and for the two-point velocity correlation function,
reveal strong intermittency for high-order moments of velocity differences.

By adopting the hypothesis of existence of the operator product expansion
in the limit of small but nonzero viscosity, $\nu \rightarrow +0$,
Polyakov \cite{AMP} reduced the problem of calculation of high-order moments
to an exactly solvable quantum mechanics and qualitatively explained the
results of numerical simulations.

Another approach was based on the equivalence of the stochastic Burgers
equation, Eqs.(\ref{burgers},\ref{covariance}), with the Martin-Siggia-Rose
field theory. Employing the saddle-point 
approximation in the corresponding path integral, Eq. (\ref{pathintegral}), 
Gurarie and Migdal 
\cite{GM} found the same generating functional of the 
velocity correlation functions as was predicted by Polyakov. 
The classical solution that provides minimum of the action was named the 
"instanton".

The purpose of this Letter is to show that the Burgers equation driven by
an external random force, Eq.~(\ref{burgers}), possesses two infinite
symmetry groups that are broken by the Gaussian distribution of a pumping
force, Eq.~(\ref{covariance}), thus resulting in the instanton solution.
The exact integration over the volumes of the symmetry group corresponds to
the description of fluctuations around the instanton and leads finally to
the Polyakov's exactly solvable quantum mechanics.

{\it Martin-Siggia-Rose formalizm.}---Let us first consider the 
functional Langevin equation of the general type
\begin{equation}
\partial_t {u} + {\cal F}(x,[{u}]) = {f}(x,t), \label{langevin}
\end{equation}
here ${f}(x,t)$ is a functional Gaussian process with the covariance
operator K. Local functional ${\cal F}(x,[{u}])$ completely characterizes the
dynamics of the system. 

Then, the Martin-Siggia-Rose formalizm
permits one to obtain a formal path integral representation for the expectation
values of the observables ${\cal O}[{u}]$ in the stationary state
\cite{MSR,JP,DP,GK,JZJ}
\begin{eqnarray}
\lefteqn{\langle {\cal O}[{u}] \rangle=\int {\cal D}{f}~{\cal O}[{u}]~
\delta(\partial_t u + {\cal F} - {f})
e^{-(f,K^{-1}f)/2}}\\
&&=\int {\cal D}\mu{\cal D}{f}~
{\cal O}[{u}]~\delta(\partial_t u + {\cal F} - {f})
e^{-(\mu,K\mu)/2+i(\mu,f)},\nonumber
\end{eqnarray}
where, we have introduced an auxiliary field ${\mu}$ to rewrite Gaussian path 
integral over the field $f$. Now, let us pass in this path integral from the 
integration over the force field ${\cal D}f$ to the integration over the 
velocity field ${\cal D}u$. Using the $\delta$-function we obtain
\begin{equation}
\langle {\cal O}[{u}] \rangle=
\int {\cal D}{\mu}{\cal D}{u}~{\cal O}[{u}]~{J}[u]~
\exp\left(-S[{u},{\mu}]\right).
\label{msrintegral}
\end{equation}
The action of the effective field theory can be
written as
\begin{equation}
S[{u},{\mu}]= -i ({\mu},\partial_t {u} + {\cal F})
+ \frac{1}{2}({\mu}, K {\mu}).
\end{equation}
The functional ${J}[u]$ is the Jacobian of the force-to-velocity
transformation 
\begin{equation}
{J}[u]=\det\left|\!\left|\frac{\delta f}{\delta u}\right|\!\right|
=\det\left|\!\left|\partial_t+\frac{\delta{\cal F}}{\delta u}\right|\!\right|.
\label{determinant}
\end{equation}
Here, the variational derivative $\delta{\cal F}/\delta {u}$
is determined by the usual equality
\begin{equation}
{\cal F}(x,[{u}+\delta{u}])-{\cal F}(x,[{u}])=\int
\frac{\delta{\cal F}(x,[{u}])}{\delta {u}(y)}~\delta{u}(y)~dy.
\label{var}
\end{equation}
Formal calculation of the determinant, Eq. (\ref{determinant}), leads to the result
\begin{equation}
{J}[u]=\exp \left(\theta(0)~{\rm Tr}~\frac{\delta{\cal F}}{\delta {u}}\right),
\label{JMSR}
\end{equation}
involving ill-defined quantity $\theta(0)$, where $\theta(t)$ is usual step
function. To give a meaning to this expression one can make use of the
usual limiting procedure in which time interval $(t_i,t_f)$ is subdivided
into $N$ equal subintervals of duration $\Delta t$ each, and the 
differential relation inside is replaced by a difference one \cite{JZJ}. 
After taking the continuous limit one finds: $\theta(0)=1/2$.

In the perturbative treatment of the path integral this Jacobian
term is irrelevant because it can always be absorbed by the proper 
redifinition of the Green functions. Nevertheless, perfoming
manipulations with the path integral we should take care about this
Jacobian term. For stochastic Burgers equation the local functional 
${\cal F}(x,[u])$ can be defined as follows:
\begin{equation}
{\cal F}(x,[u])=\int dy~\delta(x-y)~
\left(u u_y - \nu u_{yy}\right).
\end{equation}
Its variational derivative, Eq. (\ref{var}), is then equal to
\begin{eqnarray}
\frac{\delta{\cal F}(x,[u])}{\delta {u}(y)}&=&
\delta'(x-y) u-\nu\delta''(x-y),
\end{eqnarray}
and the trace term formally becomes 
\begin{eqnarray}
{\rm Tr} \frac{\delta{\cal F}}{\delta {u}}&=&
\int dtdx\frac{\delta{\cal F}(x,[u])}{\delta {u}(x)}\\
&=&\delta'(0)\int dtdx~u-\nu\delta''(0)\int dtdx.\nonumber
\end{eqnarray}
Here, the first term vanishes for symmetry
reasons and the second term does not depend on $u$ and 
can be absorbed by the normalization factor in the path integral. 
Hence, for the Burgers equation the Jacobian of the force-to-velocity 
transformation is equal to some irrelevant constant.

So, the generating functional of the velocity correlation functions
\begin{equation}
\langle F[\lambda] \rangle =
\left\langle\exp\int dx~\lambda(x)u(x,0)\right\rangle,~~~\int dx~\lambda(x)=0,
\label{correlation}
\end{equation}
can be represented by the path integral
\begin{equation}
\langle F[\lambda] \rangle=\int {\cal D}\mu{\cal D}u~F[\lambda]~
\exp\left(-S[ u ,\mu]\right),
\label{pathintegral}
\end{equation}
where action of the effective field theory is equal to
\begin{eqnarray}
\lefteqn{S[u,\mu]=-i\int dtdx\mu(u_t+uu_x-\nu u_{xx})}
\label{action}\\
& &~~~~~~~~~~~+ \frac{1}{2}\int dtdxdx' \mu(x,t)\kappa(x-x')\mu(x',t),
\nonumber
\end{eqnarray}
and time integration runs over the interval $(t_i,t_f)$. 
For the sake of simplicity, we will consider the time interval $(-\infty,0)$ 
although all the following does not depend on the particular choice.
The covariance $\kappa(x)$ is supposed to be a slowly varying even function 
with the 
expansion
\begin{equation}
\kappa(x) = \kappa(0) - \frac{\kappa_0}{2}~x^2,~~~|x|\ll\sqrt{\frac{\kappa(0)}{\kappa_0}}\equiv l,
\label{kernel}
\end{equation}
and quickly turns into zero when $|x|\gg l$. The interval $l$ characterises
the correlation length of the random force and we only study 
velocity correlation functions within this interval.

{\it Symmetries of the Burgers equation.}---We start our consideration with a
quite natural question: what is the most general transformation leaving
invariant the Burgers equation driven by an external force, Eq.~(\ref{burgers}),
or, in other words, given two functions $u(x,t)$ and $f(x,t)$ satisfying
the Burgers equation, what is the most general transformation that produces
another pair of functions satisfying the same equation?

Making use of the methods of Lie group analysis of differential equations
\cite{NHI} it is possible to prove that all such transformations form the
symmetry group which consists of the two infinite subgroups: generalized
Galilean transformations (G),
\begin{mathletters}
\begin{eqnarray}
\check{t} &=& t, \label{G1}\\
\check{x} &=& x + a(t), \label{G2}\\
\check{u} &=& u + a'(t),\label{G3} \\
\check{f} &=& f + a''(t),\label{G4}
\end{eqnarray}
\label{G}
\end{mathletters}
and time reparametrizations (L),
\begin{mathletters}
\begin{eqnarray}
\tilde{t} &=& b(t), \label{L1}\\
\tilde{x} &=& x\sqrt{b'(t)}, \label{L2}\\
\tilde{u} &=& \left( u + \frac{x}{2}\frac{b''(t)}{b'(t)} \right)/\sqrt{b'(t)}, \label{L3}\\
\tilde{f} &=& \left( f + \frac{x}{2} \{b(t),t\}\right)/\sqrt{b'(t)^3}, \label{L4}
\end{eqnarray}
\label{L}
\end{mathletters}
where $a(t)$ is an arbitrary function; $b(t)$ maps the time interval
$(t_i,t_f)$ onto itself and $b'(t)>0$. The braces
\begin{equation}
\{b(t),t\}=\frac{b'''}{b'}-\frac{3}{2} \left( \frac{b''}{b'} \right) ^2
\end{equation}
are known in the theory of complex functions as the "Schwarzian derivative".

Symmetry transformations (\ref{G},\ref{L}), although not depending
explicitly on the viscosity $\nu$, are stipulated by the structure of the
diffusion term. In particular, these transformations leave also invariant,
along with the Burgers equation itself,  the following relation,
\begin{equation}
(dx-udt)^2 \sim dt, \label{diffusion}
\end{equation}
which coincides (up to convective term, $udt$) with a similar relation
for a diffusion process.

The symmetry groups corresponding to the transformations
(\ref{G},\ref{L}) are generated by the infinitesimal operators
\begin{mathletters}
\begin{eqnarray}
G(\alpha(t))&=&\alpha\frac{\partial}{\partial x}+
             \alpha'\frac{\partial}{\partial u}+
             \alpha''\frac{\partial}{\partial f},\\
L(\beta(t))&=&\beta\frac{\partial}{\partial t}+
          \frac{1}{2}\beta' x \frac{\partial}{\partial x} +
          \frac{1}{2}\left(\beta'' x - \beta' u \right) \frac{\partial}{\partial u}\nonumber\\
&&+\frac{1}{2}\left(\beta''' x - 3\beta'' f \right) \frac{\partial}{\partial f} \label{Linf},
\end{eqnarray}
\end{mathletters}
which, in the basis
\begin{equation}
G_r = G(t^{r+1/2}),~~~L_n = L(t^{n+1}),
\end{equation}
form Lie algebra with the commutation relations
\begin{mathletters}
\begin{eqnarray}
\left[G_r,G_s\right] &=& 0,\\
\left[L_n,G_s\right] &=& (s-n/2)~G_{n+s},\\
\left[L_n,L_m\right] &=& (m-n)~L_{n+m},
\end{eqnarray}
\label{algebra}
\end{mathletters}
here $n,m$ are integers and $r,s$ are half-integers.

{\it Action transformation laws.}---Up until now, we have not been concerned
about the probability distribution of the pumping force. However, if the force
$f(x,t)$ is a Gaussian random function, then its probability measure is not
invariant under the symmetry transformations. It is easy to check
directly that the action (\ref{action}) which in fact defines such a probability
measure changes under the generalized Galilean transformations as
\begin{equation}
\check{S}[\check{u},\check{\mu}] = S[u,\mu] - i \int dtdx~\mu(x,t)~a''(t),
\label{tildeSG}
\end{equation}
and under the time reparametrizations as
\begin{eqnarray}
\lefteqn{\tilde{S}[\tilde{u},\tilde{\mu}] = -i\int dtdx\mu(u_t+uu_x-\nu u_{xx})}\nonumber \\
&&~~~~+ \frac{1}{2}\int dtdxdx'~b'^2~\mu(x,t)\kappa(\sqrt{b'}(x-x'))\mu(x',t)\nonumber\\
&&~~~~-\frac{i}{2} \int dtdx~x\mu(x,t)~\{b(t),t\}.
\label{tildeSL}
\end{eqnarray}
Hence, we come to the conclusion that it is the Gaussian distribution
of the pumping force which breaks down the infinite symmetry group of the
Burgers equation (\ref{burgers}).

The action is still invariant under the finite subgroup which consists
of spatial translations generated by the infinitesimal operator $G_{-1/2}$,
Galilean transformations generated by the operator $G_{1/2}$, and
time translations with the generator $L_{-1}$. According to the Noether
theorem \cite{NHI} there are three conservation laws corresponding to the
three-parameter subgroup of variational symmetries:
the momentum conservation
\begin{equation}
P = -i\int dx~\mu~u_x, \label{P}
\end{equation}
the conservation of the center-of-mass motion
\begin{equation}
M = i\int dx~\mu~+tP, \label{M}
\end{equation}
and the energy conservation
\begin{equation}
E = i\int dx~\mu~u_t + L, \label{E}
\end{equation}
here $L$ is the Lagrangian $S = \int dt L$.

{\it The Faddeev-Popov method.}---If the action were invariant
with respect to the infinite symmetry group, we could employ the
Faddeev-Popov method to eliminate the corresponding degrees of freedom
\cite{JZJ}. As we will see later, in our case the same method separates
the degrees of freedom of the symmetry group transformations and leads
finally to the exactly solvable quantum mechanics for the separated modes.
The Faddeev-Popov method consists of the three steps. At first, considering
the field theory
\begin{equation}
Z=\int {\cal D}A \exp(-S[A]),
\label{integral}
\end{equation}
with the action, $S[A]$,
invariant under the group ${\cal G}$ of gauge transformations
$S[A^g]=S[A]$, we define the equation $f(A)=0$ which fixes uniquely the
gauge degrees of freedom; i.e., the equation $f(A^g)=0$ should provide the
unique solution for the group element $g$ for an arbitrary chosen field
configuration $A$. Then, we define the functional Jacobian $J_{\cal G}[A]$
by the condition
\begin{equation}
J_{\cal G}[A]\int{\cal D}g~\delta\left(f(A^g)\right) = 1,
\label{identity}
\end{equation}
where $J_{\cal G}[A]=\det\|{\delta f(A^g)}/{\delta g}\|$, and integration
runs over the volume of the group ${\cal G}$. It is clear that the
functional $J_{\cal G}[A]$ is invariant by construction under the gauge
transformations, $J_{\cal G}[A^g]=J_{\cal G}[A]$. So, we can calculate the
Jacobian only for the identity element of the group. Finally, if we insert
the identity (\ref{identity}) into the functional integral (\ref{integral})
and shift the variables $A^g\rightarrow A$, we separate the degrees of
freedom corresponding to the group volume
\begin{equation}
Z=\int {\cal D} A ~J_{\cal G}[A]~\delta(f(A))~\exp(-S[A])\int {\cal D} g.
\end{equation}
Now, we can apply the same method to integrate over the volume of the
groups of generalized Galilean transformations and time reparametrizations.

{\it Integration over the volume of the group $G$}.---The gauge fixing equation in this
case is: $u(0,t)=0$, and we consider the identity
\begin{equation}
J_G[u]\int{\cal D}a(t)~\delta\left(a'-u(a,t)\right) = 1,\label{idG}
\end{equation}
which is the definition of the Jacobian $J_G[u]$; integration runs
over the volume of the group $G$. The explicit form of the Jacobian, $J_G$,
is given by the functional
determinant
\begin{equation}
J_G[u]=\det\left|\!\left|\partial_t-\frac{\delta u(a(t),t)}{\delta a(t)}\right|\!\right|.
\end{equation}
For the identity element of the group, $a(t)=0$, this expression takes the form
\begin{equation}
J_G[u]=\det\|\partial_t-u_x(0,t)\|.
\end{equation}
To preserve the causality of the Langevin equation after the generalized
Galilean transformation we should propagate operator, whose determinant
we are calculating, backward in time.
Standard calculation \cite{JZJ} then gives the result
\begin{equation}
J_G[u]=\exp\left((1-\theta(0))\int dt~u_x(0,t)\right),
\label{JG}
\end{equation}
involving the same ill-defined quantity $\theta(0)$ that appeared 
in the derivation of the Martin-Siggia-Rose field theory,
Eq. (\ref{JMSR}). As we will see later the choice, $\theta(0)=1/2$, is
self-consistent in the sense that it ensures finally the existence
of the steady state. Actually, the value of $\theta(0)$ plays exactly the
same role as the {\it B-anomaly} term in the Polyakov operator product 
expansion. 

Now, we insert the identity (\ref{idG}) into the functional integral
(\ref{correlation}). Then, after the properly defined generalized Galilean
transformation (\ref{G}), we turn the argument of the $\delta$-function into
the gauge fixing term; i.e., turn it into the equation of the surface in the
functional space that intersects the orbits of the symmetry group only once.
Finally, we get for the generating functional of the velocity correlation
functions (\ref{correlation})
\begin{eqnarray}
\langle F[\lambda]\rangle&=&\int{\cal D}\mu\int{\cal D}a~\exp(-S_G[a,\mu])\nonumber\\
&&\int{\cal D}u~\delta(u(0))~J_G~F[\lambda]~\exp(-S[u,\mu]),
\label{corr}
\end{eqnarray}
here we used the condition, $\int dx~\lambda(x)=0$, which ensures the 
invariance
of the functional $F[\lambda]$ under the generalized Galilean transformations;
\begin{equation}
S_G[a,\mu] = - i \int dtdx~\mu(x,t)~a''(t)
\end{equation}
is the effective action of the corresponding modes. In fact, all these modes
can be integrated out, leading to the additional constraint
\begin{equation}
\int{\cal D}a(t)\exp(-S_G[a,\mu])= \delta\left(\partial_t^2\int dx\mu\right).
\end{equation} From this it follows that the zero moment of the auxiliary
field $\mu$ is
just a linear function of time
\begin{equation}
\pi_0=\int dx\mu=i(Pt-M),
\end{equation}
where $P$ and $M$ are the integrals of motion (\ref{P},\ref{M}).
In the frame of reference moving together with the center of mass of the
Burgers fluid both of the integrals are equal to zero and $\pi_0\equiv 0$.

The integration over the volume of the group of generalized Galilean
transformations has first been proposed in Refs.~\cite{GM,BFKL} to
justify the stability of the saddle-point approximation in the path
integral.

{\it Integration over the volume of the group $L$}.---The gauge fixing equation is:
$u(x_0,t)=0$, where $x_0\ne 0$, and we consider the identity
\begin{equation}
J_L[u]\int{\cal D}b(t)~\delta\left(\frac{x_0}{2}\frac{b''}{b'}-u(x_0\sqrt{b'},b)~\sqrt{b'}\right) = 1,
\label{idL}
\end{equation}
The explicit expression for the Jacobian $J_L$ is
\begin{equation}
J_L[u]=\det\left|\!\left|\frac{\delta}{\delta b(t)}\left(\frac{x_0}{2}
\frac{b''}{b'}-u(x_0\sqrt{b'},b)~\sqrt{b'}\right)\right|\!\right|,
\end{equation}
which in the new variables
\begin{equation}
\ln \sqrt{b'(t)}=\sigma(b),~~~\frac{b''(t)}{2b'(t)^2}=\sigma'(b),
\end{equation}
can be rewritten as
\begin{equation}
J_L[u]=\det\left|\!\left|
\frac{\delta
\left\{x_0\sigma'-u(x_0~e^\sigma,b)e^{-\sigma}\right\}e^{2\sigma}
}{\delta\sigma}\right|\!\right|\det\left|\!\left|\frac{\delta\sigma}{\delta b}\right|\!\right|.
\end{equation}
Again, due to invariance of the Jacobian under the group of time
reparametrizations it can be computed only for the identity element of the
group
\begin{equation}
J_L[u]=\det\|\partial_t-u_{x}(x_0,t)\|,
\label{det}
\end{equation}
where the unessential constant coefficient that does not depend on $u$ is
omitted and constraint, $u(x_0,t)=0$, is used. Calculating the determinant
along the lines indicated above  we get
\begin{equation}
J_L[u]=\exp\left(\frac{1}{2}\int dt~ u_{x}(x_0,t)\right).
\label{J_L}
\end{equation}
Let us now insert the identity ($\ref{idL}$) into the functional integral
(\ref{corr}). Then, after a properly defined time reparametrization, the
generating functional of the velocity correlation functions is converted
into
\begin{eqnarray}
\lefteqn{\langle F[\lambda]\rangle=}\nonumber\\
&&\int\frac{{\cal D}\mu}{{\cal D}\pi_0}{\cal D}u
~\delta(u(0))~\delta(u(x_0))~J_G~J_L~F[\lambda]~\exp{(-S_0)}\nonumber\\
&&\int {\cal D}b~\exp\left\{-S_L[b,\pi]+
\frac{b''(0)}{2b'(0)^2}\int dx~x\lambda(x)
\right\},
\label{F_L}
\end{eqnarray}
where $\pi(t)=\int dx~x\mu(x,t)$ and
\begin{equation}
S_0[u,\mu] =  -i\int dtdx\mu(u_t+uu_x-\nu u_{xx}),
\end{equation}
is the action of the modes remaining. Due to the special choice of the
noise covariance, Eq.(\ref{kernel}), it does not depend on
the pumping force.

In the limit of the small but nonzero viscosity, $\nu \rightarrow +0$, we
can separate also the integration over $\pi$.
The desired effective action of
the separated modes $b(t)$ and $\pi(t)$,
\begin{eqnarray}
S_L[b,\pi] &=& - \frac{i}{2} \int dt~\pi(t)~\{b(t),t\} -
\frac{1}{2}\int dt~ \frac{b''(t)}{2b'(t)}\nonumber \\
&& + \frac{\kappa_0}{2}\int dt~b'(t)^3~\pi(t)^2,
\label{S_L}
\end{eqnarray}
follows then from the action transformation law under the time
reparametrizations, Eq.~(\ref{tildeSL}), and the analogous transformation law
for the Jacobian
\begin{equation}
\tilde{J}_G[\tilde{u}] = J_G[u] \exp\left(\frac{1}{2}\int dt~ \frac{b''}{2b'}\right).
\end{equation}

Separating the modes, we obtain  another ambiguous parameter. Namely, we
can choose an arbitrary normalization factor, $Z_0$, for the partition 
function 
of the modes of the symmetry group transformations and the inverse one,
$1/Z_0$, for the partition function of the modes remaining. This parameter
plays exactly the same role as the {\it A-anomaly} term in Polyakov 
theory. Again, the choice $Z_0=1$ is self-consistent and it 
ensures the existence of the steady state (see also \cite{SB}).

It should be stressed at this point that the gauge fixing term
$\delta(u(x_0))$ works properly only if the degrees of freedom associated
with the field $a(t)$ have already been integrated out; i.e., we cannot
change the order and integrate first over the volume of the group $L$.

{\it Equivalent Quantum Mechanics.}---The action (\ref{S_L}) is equivalent
to the Polyakov's exactly solvable quantum mechanics. To make it obvious,
let us pass from the variables $t$, $b(t)$ and $\pi(t)$ to the new ones
\begin{equation}
b=b(t),~~q(b)=\frac{b''(t)}{2b'(t)^2},~~p(b)=b'(t)\pi(t),
\end{equation}
and finally get
\begin{equation}
S_L[q,p] = \int db~\left\{-ip~(q'+q^2)-\frac{3}{2}~q+\frac{\kappa_0}{2}~p^2\right\},
\end{equation}
where the additional term, $-q$, comes from the Jacobian ${\cal D}\pi{\cal D}b/{\cal D}p{\cal D}q$.
This action corresponds to the quantum mechanics (in imaginary time) with the
Hamiltonian \cite{JZJ}
\begin{equation}
{\bf H}=\frac{\kappa_0}{2}\left({\bf p}-\frac{i~q^2}{\kappa_0}\right)^2
+\frac{q^4}{2\kappa_0}-\frac{3q}{2}.
\end{equation}
Following Polyakov \cite{AMP}, we can find the zero energy eigenfunction
of this Hamiltonian and finally calculate the generating functional of the
velocity correlation functions,
\begin{equation}
\langle F[\lambda]\rangle =
\exp \left\{ \frac{\sqrt{2\kappa_0}}{3}\left[\int dx~x\lambda(x)\right]^{3/2}\right\}.
\end{equation}

In conclusion I would like to mention that the method proposed in this Letter
seems to have a wider applicability than just to the Burgers equation
considered. Any stochastic equation of the Langevin type that possesses
infinite symmetry group can be treated this way.

I am grateful to V.B.~Priezzhev, V.~Gurarie, S.~Boldyrev and E.~Balkovsky for 
fruitful discussions. I would like also to thank T. Spencer for kind
hospitality during my visit to Institute for Advanced Study, Princeton.
This work was supported by the Russian Foundation for 
Basic Research through Grant No. 95-01-00257 and by the International Center 
for Fundamental Physics in Moscow through INTAS Grant No. 93-2492.

\end{document}